\documentclass[pra,twocolumn,nofootinbib,floatfix,10pt]{revtex4-1}
\pdfoutput=1 
\usepackage[utf8]{inputenc}

\usepackage{graphicx}
\usepackage{amsmath, amsxtra, amssymb, mathtools, isomath, nccmath, xspace}
\usepackage{ wasysym }
\usepackage{setspace, changepage}
\usepackage{bbold}
\usepackage{hhline}
\usepackage{multirow}
\usepackage{hyperref}
\usepackage{helvet}

\newcommand{\ket}[1]{\ensuremath{\left| #1 \right\rangle}\xspace}

\long\def\symbolfootnote[#1]#2{\begingroup%
\def\thefootnote{\fnsymbol{footnote}}\footnotetext[#1]{#2}\endgroup}

\date{\today}

\begin{document}

\title{  
Time-Resolved Observation of Spin-Charge Deconfinement in Fermionic Hubbard Chains
}

\author{Jayadev~Vijayan$^{1,2,\ast,\dag}$}%
\author{Pimonpan~Sompet$^{1,2,\ast}$}%
\author{Guillaume Salomon$^{1,2}$}%
\author{Joannis Koepsell$^{1,2}$}%
\author{Sarah Hirthe$^{1,2}$}%
\author{Annabelle Bohrdt$^{2,3}$}%
\author{Fabian Grusdt$^{2,3}$}%
\author{Immanuel Bloch$^{1,2,4}$}%
\author{Christian Gross$^{1,2}$}%

\affiliation{$^{1}$Max-Planck-Institut f\"{u}r Quantenoptik, 85748 Garching, Germany}
\affiliation{$^{2}$Munich Center for Quantum Science and Technology (MCQST), Schellingstra{\ss}e 4, 80799 M{\"u}nchen, Germany}
\affiliation{$^{3}$Department of Physics and Institute for Advanced Study, Technical University of Munich, 85748 Garching, Germany}
\affiliation{$^{4}$Fakult\"{a}t f\"{u}r Physik, Ludwig-Maximilians-Universit\"{a}t, 80799 M\"{u}nchen, Germany}

\symbolfootnote[1]{These authors contributed equally to this work}
\symbolfootnote[2]{Electronic address: {jayadev.vijayan@mpq.mpg.de}}

\begin{abstract}
Elementary particles such as the electron carry several quantum numbers, for example, charge and spin.
However, in an ensemble of strongly interacting particles, the emerging degrees of freedom can fundamentally differ from those of the individual constituents. 
Paradigmatic examples of this phenomenon are one-dimensional systems described by independent quasiparticles carrying either spin (spinon) or charge (holon).
Here we report on the dynamical deconfinement of spin and charge excitations in real space following the removal of a particle in Fermi-Hubbard chains of ultracold atoms.
Using space- and time-resolved quantum gas microscopy, we track the evolution of the excitations through their signatures in spin and charge correlations.
By evaluating multi-point correlators, we quantify the spatial separation of the excitations in the context of fractionalization into single spinons and holons at finite temperatures. 
\end{abstract}
\maketitle

\subsubsection*{\textbf{\emph{Introduction}}}

Strongly correlated quantum systems are known to show peculiar behaviour, which often cannot be attributed microscopically to the properties of weakly dressed individual electrons such as in ordinary metals.
Instead, the collective nature of the excitations can lead to the emergence of new quasiparticles, which are fundamentally distinct from free electrons.
This behaviour is a hallmark of one-dimensional (1D) quantum systems, where electron-like excitations do not exist, but are replaced by decoupled collective spin and charge modes~\cite{Giamarchi2004}.
These two independent excitation branches feature different propagation velocities~\cite{Voit1995} and have previously been studied in the Luttinger liquid regime of quasi-1D solids using spectroscopic techniques, such as angle-resolved photoemission spectroscopy~\cite{Kim1996,Segovia1999,Kim2006} and conductance measurements in metallic quantum wires~\cite{Auslaender2005,Jompol2009,Tserkovnyak2003}.
Cold atom experiments have been used extensively to study attractive 1D bosonic and fermionic gases~\cite{Kinoshita2004,Paredes2004,Haller2009,Jacqmin2011,Fabbri2015}, but the investigation of repulsive 1D fermionic gases has been more recent~\cite{Yang2018,Boll2016}. 
Such experiments can probe the regime in between the low-energy Luttinger liquid description and the spin-incoherent Luttinger liquid for which the temperature is on the order of or exceeds the magnetic energy~\cite{Fiete2007}.
Recent equilibrium signatures of spin-charge separation have been observed in ultracold lattice gases using quantum gas microscopy~\cite{Hilker2017,Salomon2019}.
However, a real-space tracking of the dynamics of the individual excitations signalling their deconfinement has been lacking so far.

Here we demonstrate dynamical spin-charge separation directly by performing a local quench in a 1D gas of ultracold fermionic atoms and subsequently monitoring the evolution of the system with spin- and density-resolved quantum gas microscopy~\cite{Boll2016} (see Fig. 1).
The local quench is realized by the high fidelity removal of one atom from a single site of a 1D optical lattice initially filled with nearly one atom per site and short-range antiferromagnetic spin correlations~\cite{Cheuk2016,Parsons2016,Boll2016,Brown2017}.
In the subsequent dynamics, we observe the emergence of two apparently independent excitations propagating at different velocities~\cite{Recati2003,Kollath2005,Kollath2006}, which we assign to spinons and holons based on their characteristic signatures in the spin and charge (density) sectors.

Through the evaluation of multi-point correlators, such as the spin correlations across the propagating hole and the magnetization fluctuations in a finite region of space, we quantify the spatial separation of the excitations. 
We find the measured correlations to agree well with the expected signatures of fractionalization at finite temperatures.

\begin{figure*}[t!]
\centering
\includegraphics[width=0.8\textwidth]{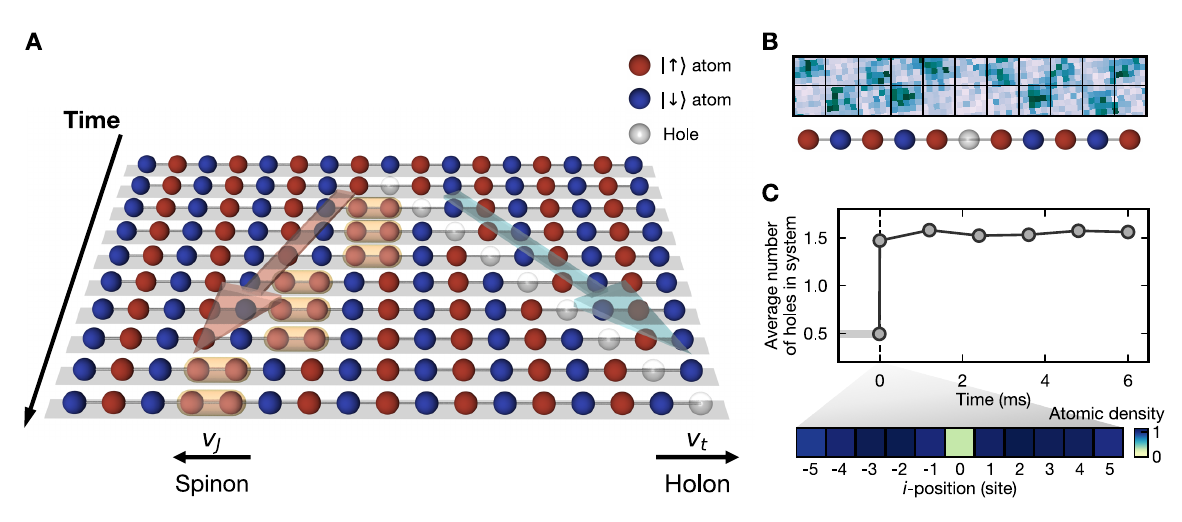}
\caption{
\textbf{Probing spin-charge deconfinement with cold atoms}.
\textbf{A,} Cartoon depicting fractionalization of a fermionic excitation into quasiparticles.
The dynamics is initiated by removing a fermion from the Hubbard chain.
This quench creates a spin (spinon) and a charge (holon) excitation, which propagate along the chain at different velocities $v_{J}$ and $v_{t}$.
\textbf{B,} Using quantum gas microscopy, we simultaneously detect the spin and density on every site of the chain after a variable time after the quench.
\textbf{C,} 
Average number of holes in the chain as a function of time. Error bars denote 1 s.e.m. The quench, performed at {$0$ ms}  creates a hole with a probability of $\sim 78\%$ in the central site of the chain (bottom).
}

\end{figure*}

\subsubsection*{\textbf{\emph{State preparation}}}

Our experiment~\cite{Salomon2019} starts by loading a two-dimensional balanced spin mixture of \textsuperscript{6}Li atoms in the lowest two hyperfine states into several 1D tubes using an optical lattice of spacing $a_y = 2.3\,\mu$m along the $y$-direction.
Next, a lattice of spacing $a_x = 1.15\,\mu$m is ramped up along the $x$-direction.
By varying the hopping strength along the $x-$direction from $t/h = 190\,$Hz to $t/h = 410\,$Hz, we realize Fermi-Hubbard chains with $U/t \sim 8 - 20$, where $U$ is the onsite interaction energy.
We fix the total atom number in the gas to around $75$ by the choice of the evaporative cooling parameters such that the resulting Hubbard chains are prepared close to half-filling in the center of the harmonically confined cloud.
This produces at least three 1D chains of mean length $13$ atoms, each with a unity filled region of about 9 sites.
To perform a local quench in which a single atom is simultaneously removed from all chains, we use an elliptically shaped near-resonant laser beam at $671\,$nm focussed to a waist of $\sim 0.5\,\mu$m along its narrow direction.
This pushout beam is pulsed on for $20\,\mu$s addressing the central sites.
The power and alignment of the pushout beam is adjusted such that the probability of spin-independent removal of an atom from the addressed site is $\sim 78\%$, with $\sim 14\%$ chance of affecting the nearest neighbouring sites (see Supplementary Information).
After the quench, we let the system evolve for a variable hold-time before imaging the spin and density distributions.
To collect statistics, the experiment is repeated several thousand times for a given evolution time.  
 
\subsubsection*{\textbf{\emph{Holon and spinon dynamics}}}

We first investigate the difference in the dynamics of holons and spinons by preparing 1D Hubbard chains with $t = h \times 250\,$Hz and $U/t = 15$, corresponding to an exchange interaction of $J = h  \times 65\,$Hz, and then performing the local quench.
A natural observable to characterize the subsequent dynamics of holons is the spatially resolved hole density distribution $\langle \hat{n}^h_{i}\rangle$ in each chain, where $i$ labels the lattice sites.
The observed distribution broadens as a function of time with a light-cone-like ballistic propagation of the wavefront (see Fig. 2A).
It starts from the addressed site and reaches the edge of the unity-filled region of the chain in $5\,\tau_{t}$, where $\tau_{t} = h\times(4\pi t)^{-1} = 0.32$ ms is the time it takes for a hole propagating at the theoretically expected maximum group velocity $v^{\text{max}}_{t} = a_x / \tau_{t}$, to move by one site.
The coherent evolution of the hole can be seen in the evolving interference pattern of $\langle \hat{n}^h_{i}\rangle$ over time.
This dynamics is found to be in excellent agreement with a single particle quantum walk (see Fig. 2B), as expected for a spin-charge separated system.

\begin{figure*}[t!]
\centering
\includegraphics[width=0.8\textwidth]{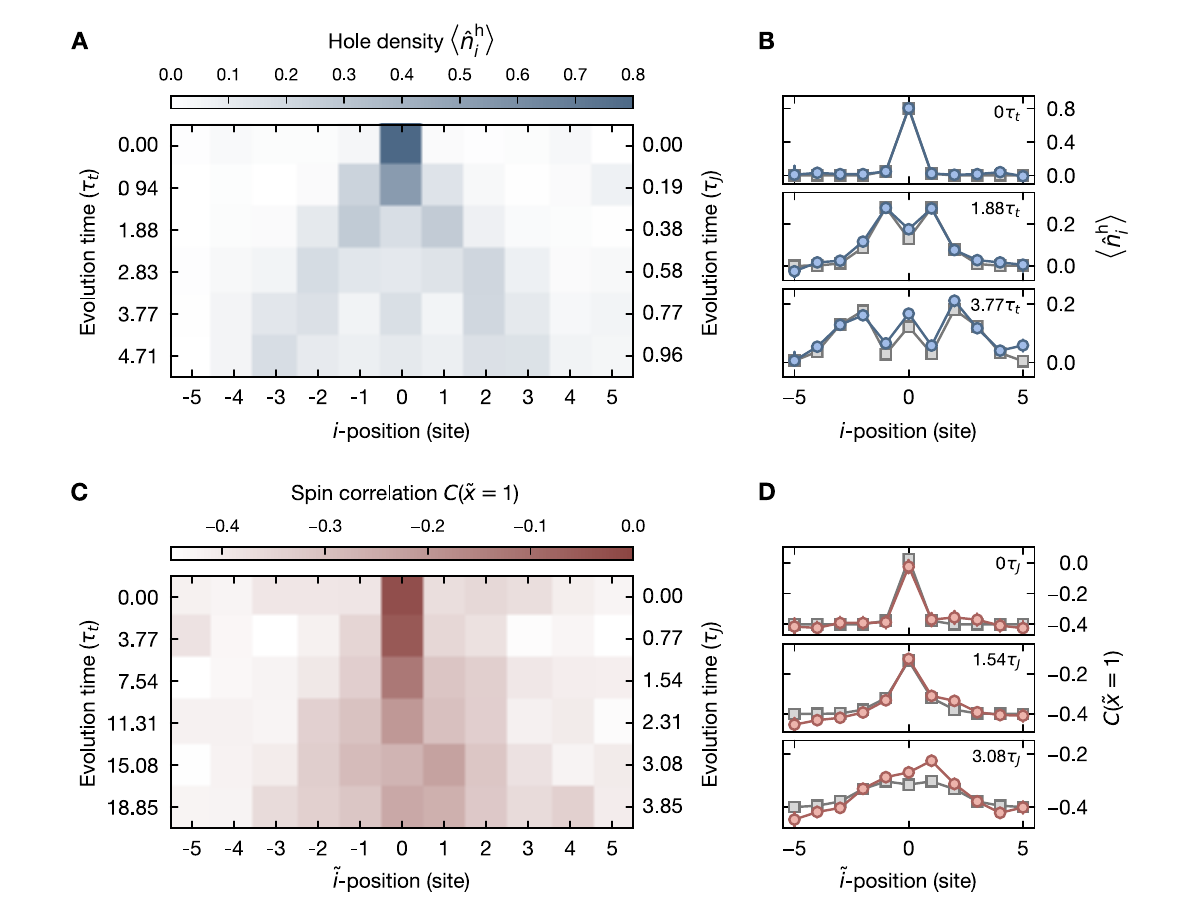}
\caption{
\textbf{Time evolution of spin and charge excitations.
} \textbf{A,} Hole density distribution $\langle \hat{n}^\text{h}_{i}\rangle$ as a function of time after the quench.
The wavefront of the distribution starts at the center of the chain and expands outwards linearly with time.
Interference peaks and dips are visible throughout the dynamics, indicating the coherent evolution of the charge excitation.
\textbf{B,} One-dimensional cuts of the experimental hole density distributions at times $0\,\tau_{t}$, $1.88\,\tau_{t}$ and $3.77\,\tau_{t}$ (blue circles) are compared with simulations of a single particle quantum walk (grey squares).
\textbf{C,} Nearest neighbour squeezed space spin correlation $C(\tilde{x}=1)$ distribution as a function of time after the quench.
\textbf{D,} One-dimensional cuts of the experimental $C(\tilde{x}=1)$ distributions at times $0\,\tau_{J}$, $1.54\,\tau_{J}$ and $3.08\,\tau_{J}$ (red circles) along with exact diagonalization simulations of the Heisenberg model (grey squares). Error bars denote 1 s.e.m.
}
\end{figure*}

\begin{figure*}[t!]
\centering
\includegraphics[width=0.8\textwidth]{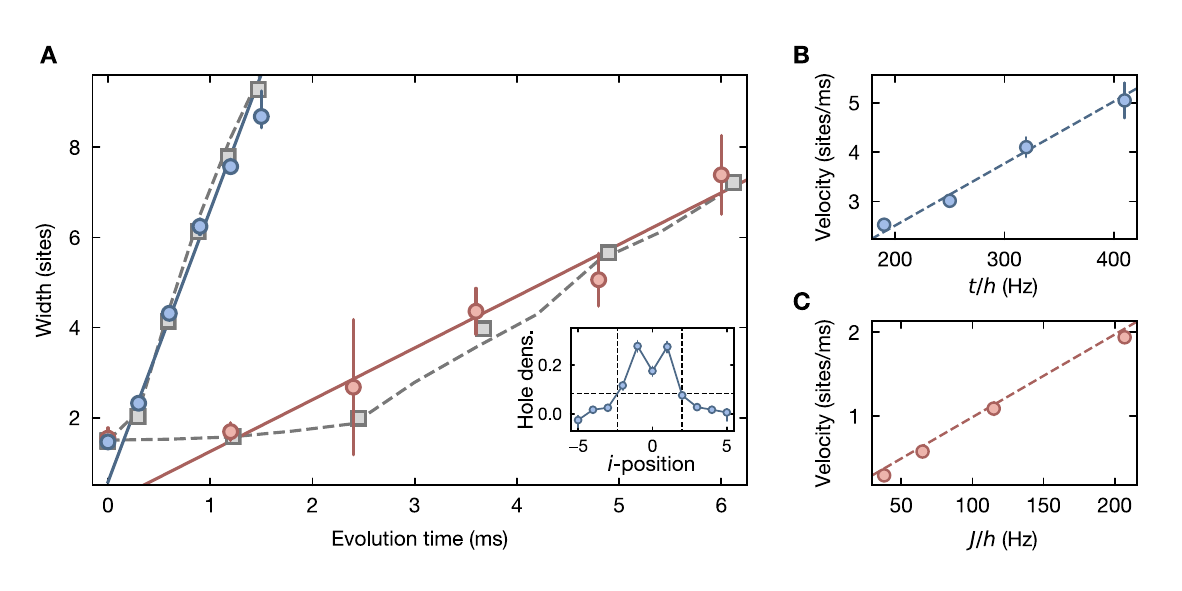}
\caption{
\textbf{Quasiparticle velocities of spinons and holons.
} \textbf{A,} Time evolution of the widths of the hole density distributions (blue circles) and nearest neighbour spin correlation distributions (red circles) after the quench.
The measured widths are defined as the full width at 30\% of maxima of the distributions (see inset).
Density and spin excitations reach the edge of the unity filled region of the chain (central 9 sites) after different evolution times.
Their dynamics are in quantitative agreement with both a single particle quantum walk for hole and exact diagonalization calculations of the Heisenberg model for the spin (grey squares). They are also found to reproduce the predictions of the extended $t-J$ model at our temperature (grey dashed lines).
The velocities of the spin ($0.58\pm0.04\,$sites/ms) and the charge ($3.08\pm0.09\,$sites/ms) excitations are obtained as half the slope of a linear fit to the data(solid blue and red lines), ignoring the width immediately after the quench.
\textbf{B,} Holon velocities as a function of $t/h$.
The velocities of the holon (blue circles) increase linearly with the tunneling rate in the chain, consistent with $v_{\text{max}}^{t} = 4\pi t a_x/h\,$ sites/ms (blue dashed line).
\textbf{C,} Spin excitation velocities as a function of $J/h$.
The velocities of the spin excitation (red circles) increase linearly with the spin-exchange coupling in the chain, consistent with $v_{\text{max}}^{J} = \pi ^{2} J a_x/h\,$ sites/ms (red dashed line). Error bars denote 1 s.e.m.
}
  
\end{figure*}

To study the time evolution of the spin excitation, we measure nearest neighbour spin correlations $C_{\tilde{i}}\,(\tilde{x}=1) = 4(\langle \hat{S}_{\tilde{i}}^{z} \hat{S}_{\tilde{i}+1}^{z}\rangle - \langle \hat{S}_{\tilde{i}}^{z}\rangle \langle \hat{S}_{\tilde{i}+1}^{z}\rangle)$ in squeezed space (denoted by $\sim$), obtained by removing holes and doublons from the chain in the analysis~\cite{Hilker2017,Salomon2019}. 
For strong interactions $U/t\,\gtrsim\,8$, the spin dynamics in squeezed space is expected to be well captured by an antiferromagnetic Heisenberg model ~\cite{Ogata1990,Zaanen2001} to which we compare our results.
We observe a strong reduction of the antiferromagnetic correlations in the direct vicinity of the quenched site immediately after the quench, demonstrating an enhanced probability to find parallel spins on neighbouring sites. 
Such a suppression is expected from the creation of spinons by the local quench (see Fig. 1A).
The region with reduced antiferromagnetic correlations spreads with time, with a light-cone-like propagation of the wavefront (see Fig. 2C).
It reaches the edge of the unity filled region in $4\,\tau_{J}$, where $\tau_{J} = h\times(\pi^{2} J)^{-1} = 1.56\,$ms is the time it takes for a spinon propagating at the theoretically expected maximum group velocity $v^{\text{max}}_{J} = a_x / \tau_{J}$, to move by one site.
In contrast to the highly coherent evolution of the hole, the finite temperature $k_{\text{B}}T/J \sim 0.75$ in our system prevents us from observing any interference effects in the spin dynamics. 
However, the observed ballistic wavefront is still expected from the Heisenberg model at our temperatures (see Supplementary Information)~\cite{Castella1995,Zotos1997}.

Next, we extract the velocities of the spin and charge excitations emerging from the quench.
We monitor the spatial width of the squeezed space correlator $C_{\tilde{i}}\,(\tilde{x} = 1)$ and hole distributions as a function of time (see Fig. 3A, inset).
We then use a linear fit to determine their respective velocities (see Fig. 3A).
For the data shown in Fig. 3A, taken at $U/t=15$, we find a ratio of $5.31\,\pm\,0.43$ between the two propagation velocities, indicating a strong difference in the velocities of the two excitation channels.
Despite the finite non-zero temperature in our system, the extracted velocities are in excellent agreement with both exact diagonalization results of an extended $t-J$ model (see Supplementary Information), as well as single particle quantum walk for a hole and a Heisenberg model prediction at our temperature for the spin excitations.\\\\
To investigate the scaling of the velocities with the tunneling and spin exchange energies $t$ and $J$, we repeat the experiment for different $U/t$ by tuning the lattice depth. 
Within our experimental uncertainties, we find the extracted velocities to be in good agreement with the maximum expected group velocities $v^{\text{max}}_{t}$ and $v^{\text{max}}_{J}$ for the two excitation channels (see Fig. 3B and 3C).
These correspond to the velocities of a free holon and spinon at the maximum group velocity allowed by their dispersion. 
Unlike the Luttinger liquid regime which only describes low energy excitations, our local quench excites all momentum modes, the fastest of which is tracked here.

\subsubsection*{\textbf{\emph{Spatial separation of the quasiparticles}}}

An essential feature of spin-charge deconfinement is the existence of unbound states of spin and charge excitations, allowing them to spatially separate over arbitrary large distances.
To quantify the dynamical deconfinement, we study the spin correlations across the propagating hole as a function of time, through the spin-hole-spin (SHS) correlator $C_{\text{SHS}}\,(2) = 4\,\langle\,\hat{S}_{i}^{z}\,\hat{n}^h_{i+1}\,\hat{S}_{i+2}^{z}\,\rangle$, a spin correlator conditioned on having a hole at site $i+1$~\cite{Hilker2017,Salomon2019}(see Fig. 4A).

\begin{figure*}[t!]
\centering
\includegraphics[width=1\textwidth]{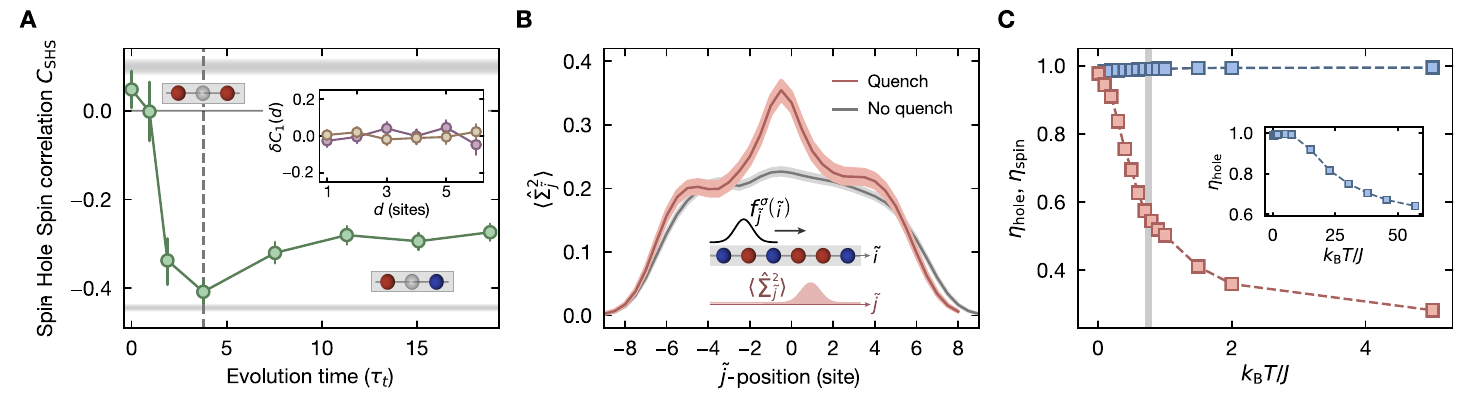}
\caption{
\textbf{Spatial deconfinement of spin and charge excitations.} 
\textbf{A,} Spin-hole-spin correlations ($C_{\text{SHS}}$) averaged over the entire chain as a function of time after the quench.
The correlator starts with a positive value consistent with the next-nearest neighbour spin correlations $C(2)$ in the absence of the quench (top grey shaded region) and turns negative, approaching the nearest neighbour spin correlations $C(1)$ without the quench (bottom grey shaded region) by $4\,\tau_{t}$.
At longer evolution times, the correlator shows reduced antiferromagnetic correlations due to the oscillating dynamics of the hole in our finite size system.
The inset shows the lack of dependence of the normalized deviation from the mean nearest neighbour correlations $\delta C_1$ on the distance $d$ from the hole at times $\sim 4\,\tau_t$ (purple) and $\sim 19\,\tau_t$ (yellow). Error bars denote 1 s.e.m.
\textbf{B,} Spatially resolved magnetization fluctuations $\langle\,\hat{\Sigma}_{\tilde{j}}^{2}\,\rangle$ in sub-regions of the chain with (red) and without (grey) the quench at $3.77 \tau_{t}$. 
The background fluctuations $\langle\,\hat{\Sigma}^{2}_{\tilde{j}}\,\rangle_{\text{BG}}$ are due to quantum and thermal fluctuations in our system.
The peak in the difference signal $\langle\,\hat{\Sigma}_{\tilde{j}}^{2}\,\rangle - \langle\,\hat{\Sigma}^{2}_{\tilde{j}}\,\rangle_{\text{BG}}$ indicates the location of the spin excitation. Grey and red shades denote 1 s.e.m without and after the quench respectively.
\textbf{C,} Efficiency of initially creating at the central site a single local spinon ${\eta_{\mathrm{spin}} = 4(\langle\,\hat{\Sigma}_{\tilde{j}=0}^{2}\,\rangle - \langle\,\hat{\Sigma}^{2}_{\tilde{j}=0}\,\rangle_{\text{BG}})}$ with $\sigma = 1.5$ sites (orange) and holon ${\eta_{\mathrm{hole}} = 1 - \langle (\hat{n}_{i=0} -1)^2\rangle}$ (blue) after an ideal quench as a function of temperature as predicted from exact diagonalization of the Heisenberg model (for the spinon) and the Hubbard model (for the holon).
With increasing temperature, $\eta_{\mathrm{spin}}$ ($\eta_{\mathrm{hole}}$, inset) decreases due to the increase of thermal spin (density) excitations, preventing the creation of a localized spinon (holon) by the quench.
Taking into account our quench efficiency, the measured amplitude is consistent with the prediction at a temperature of $k_{\text{B}}T/J = 0.75$ (grey shaded region).
}
\end{figure*}

Immediately after the quench, the hole is likely to be surrounded by parallel spins and $C_{\text{SHS}}$ retains a positive value.
The measured spin correlations are consistent with the next-nearest-neighbour correlations $C(2) = 4(\langle \hat{S}_{i}^{z}\hat{S}_{i+2}^{z}\rangle - \langle \hat{S}_{i}^{z}\rangle \langle\hat{S}_{i+2}^{z} \rangle)$ in the absence of the quench.
As the hole propagates, the sign of $C_{\text{SHS}}$ becomes negative and by $4\,\tau_t$, approaches the nearest neighbour correlations $C(1) = 4(\langle \hat{S}_{i}^{z}\hat{S}_{i+1}^{z}\rangle - \langle \hat{S}_{i}^{z}\rangle \langle\hat{S}_{i+1}^{z} \rangle)$ without the quench.
These observations indicate the decoupling of the two excitations.
At longer evolution times, the antiferromagnetic correlations across the hole are reduced.
We attribute this to the holon oscillating in the chain due to the harmonic confinement present in our system and hence to the changing overlap of the spin and charge distributions (see Supplementary Information).
The absence of binding between the spin and charge excitations beyond the immediate vicinity of the hole is shown by calculating the normalized deviation from the mean nearest neighbour correlations $\delta C_1(d) = \langle \hat{S}^z_i \hat{S}^z_{i+1}/\langle \hat{S}^z_i \hat{S}^z_{i+1}\rangle -1\rangle_{ \scalebox{0.65}{\newmoon}_{i}  \scalebox{0.65}{\newmoon}_{i+1}\scalebox{0.65}{\fullmoon}_{i+1+d\,\lor\,i-d}}$ (see Fig. 4A, inset), where $d$ is the distance of the hole from the closest of sites $i$ and $i+1$.
$\delta C_1$ shows no dependence on $d$, indicating the lack of influence of the holon on the spin excitation.

To locate the excess spin excitation in a fluctuating spinon background, we introduce an operator quantifying the local spin fluctuations in squeezed space $\hat{\Sigma}^{2}_{\tilde{j}} = (\sum_{\tilde{i}}\hat{S}^{z}_{\tilde{i}}f_{\tilde{j}}^{\sigma}(\tilde{i}))^{2}$, where $f_{\tilde{j}}^{\sigma}(\tilde{i}) = \exp(-(\tilde{i}-\tilde{j})^{2}/(2\sigma^{2}))$ is a smooth window function centered at lattice site $\tilde{j}$ with a characteristic size of $\sigma$. 
At zero temperature, this operator is expected to capture local fractional quantum numbers~\cite{Kivelson1982}. 
A single spinon located at site $\tilde{j}$, carrying a spin $1/2$, would increase $\langle\,\hat{\Sigma}_{\tilde{j}}^{2}\,\rangle$ by $1/4$, provided that the mean distance between thermal spin fluctuations is larger than $\sigma$. 

To study the spatial separation of the spin and charge excitations, we consider the chains at time $3.77\,\tau_t$, where the highest probability to detect the hole is at sites $\pm 2$. 
We post-select on chains with a single hole in the central nine sites, which is located outside the central three sites and compute $\langle\,\hat{\Sigma}^{2}_{\tilde{j}}\,\rangle$ on all sites $\tilde{j}$ for a window size $\sigma = 1.5$ (see Fig 4B).
Comparing $\langle\,\hat{\Sigma}^{2}_{\tilde{j}}\,\rangle$ with the quench to its value $\langle\,\hat{\Sigma}^{2}_{\tilde{j}}\,\rangle_{\text{BG}}$ without the quench, we observe a well localized signal extending over the central three sites, distinct from the position of the holon. 
The maximum deviation ${\langle\,\hat{\Sigma}_{\tilde{j}}^{2}\,\rangle - \langle\,\hat{\Sigma}^{2}_{\tilde{j}}\,\rangle_{\text{BG}}}$ reaches $0.13\pm0.01$, about half the value expected at zero temperature.
We attribute this difference mainly to the finite temperature of our system leading to a background density of thermal spin excitations. 
In this case, even an ideal quench would not create an initially localized spinon with unity probability and the fractionalization scenario, where a single removed particle breaks up into precisely one holon and one spinon, holds only asymptotically at zero temperature (see Fig. 4C).
A reduction in the deviation $\langle\,\hat{\Sigma}_{\tilde{j}}^{2}\,\rangle - \langle\,\hat{\Sigma}^{2}_{\tilde{j}}\,\rangle_{\text{BG}}$ from 1/4 is thus expected in our system and the measured value is in good agreement with exact diagonalization calculations of the Heisenberg model at $k_{\text{B}}T=0.75\,J$ taking into account our quench efficiency (see Supplementary Information).

\subsubsection*{\textbf{\emph{Conclusion and outlook}}}

We have demonstrated the dynamical deconfinement of spin and charge excitations in the 1D Fermi-Hubbard model by performing a local quench experiment and tracking the propagation of the quasiparticles with full spatial and temporal resolution.
The measured velocities of spin and charge excitations were found to significantly differ, leading to their spatial separation after a few tunneling times.
An interesting extension of this work would be to study spin-charge confinement dynamics in the dimensional crossover from 1D to 2D, where polaronic signatures were recently observed~\cite{Koepsell2018,Salomon2019}.
The protocol used here could directly be implemented to extract the effective mass of a polaron.
Finally, we demonstrated how quantum gas microscopy can be used to study the fate of spin-charge fractionalization at finite temperature.
This opens new perspectives to dynamically probe the doped Fermi-Hubbard model in higher dimensions and explore fractionalization in topological phases of matter.

\subsubsection*{\textbf{\emph{Acknowledgements}}}

We thank Ganapathy Baskaran, Eugene Demler, Thierry Giamarchi, Roderich Moessner and Ramamurti Shankar for useful discussions. 
We acknowledge funding by the Max Planck Society (MPG), the European Union (UQUAM Grant No. 319278, PASQuanS Grant No. 817482) and the Deutsche Forschungsgemeinschaft (DFG, German Research Foundation) under Germany’s Excellence Strategy – EXC-2111 – 390814868.
J.K. acknowledges funding from the Hector Fellow Academy and G.S. acknowledges funding from the Max Planck Harvard Research Center for Quantum Optics.
F.G. and A.B. acknowledge support from the Technical University of Munich - Institute for Advanced Study, funded by the German Excellence Initiative and the European Union FP7 under grant agreement 291763, from the DFG grant No. KN 1254/1-1, and DFG TRR80 (Project F8). A.B. also acknowledges support from the Studienstiftung des deutschen Volkes.

\subsubsection*{\textbf{\emph{Materials and Correspondence}}}
Correspondence and requests for materials should be addressed to jayadev.vijayan@mpq.mpg.de.

\newcommand{\mkch}[1]{{\color{BrickRed} #1}}
\newcommand{\mkcomm}[1]{{\color{BrickRed} MK: #1}}
\newcommand{\tbd}[1]{{\color{green} TBD #1}}
\newcommand{\mh}[1]{{\color{blue} #1}}

\newcommand{\ch}{Ch.\@\xspace}
\newcommand{\se}{Sec.\@\xspace}
\newcommand{\Se}{Sec.\@\xspace}
\newcommand{\app}{App.\@\xspace}
\newcommand{\ie}{i.e.\@\xspace}
\newcommand{\eg}{e.g.\@\xspace}

\newcommand{\mean}[1]{\langle #1 \rangle_m}
\newcommand{\ptl}{\partial}
\newcommand{\DF}[2]{\frac{d\, #1}{d\, #2}}
\newcommand{\PDF}[2]{\frac{\ptl\, #1}{\ptl\, #2}}
\newcommand{\PDFS}[2]{\frac{\ptl^2}{\ptl\, #1 \ptl\, #2}}
\newcommand{\DELF}[2]{\frac{\delta\,#1}{\delta\,#2}}
\newcommand{\ve}[1]{{\bf #1}}
\newcommand{\mat}[1]{\mathsf{#1}}
\newcommand{\nag}{{\phantom{\dagger}}}
\newcommand{\dimmu}{t^{1/2}\delta^{3/2}}
\newcommand{\den}{\rho}

\newcommand{\eqw}[1]{(\ref{#1})}
\newcommand{\eqq}[2]{Eqs.\thinspace{}(\ref{#1}) and (\ref{#2})}
\newcommand{\eqqs}[2]{Eqs.\thinspace{}(\ref{#1}) and (\ref{#2})}
\newcommand{\eqqqs}[3]{Eqs.\thinspace{}(\ref{#1}), (\ref{#2}) and (\ref{#3})}
\newcommand{\eqqqqs}[4]{Eqs.\thinspace{}(\ref{#1}), (\ref{#2}), (\ref{#3}) and (\ref{#4})}
\newcommand{\Eq}[1]{Eq.\thinspace{}(\ref{#1})}

\newcommand{\tab}[1]{Tab.\thinspace{}\ref{#1}}

\newcommand{\alg}[1]{Alg.\thinspace{}\ref{#1}}
\newcommand{\algline}[1]{\ref{#1}:}

\newcommand{\fig}[1]{Fig.\thinspace{}\ref{#1}}
\newcommand{\figg}[2]{Fig.\thinspace{}\ref{#1} and \ref{#2}}
\newcommand{\fc}[1]{({#1})}
\newcommand{\figc}[2]{Fig.\thinspace{}\ref{#1}\thinspace{}\fc{#2}}
\newcommand{\figcc}[3]{Fig.\thinspace{}\ref{#1}\thinspace{}\fc{#2} and \fc{#3}}
\newcommand{\Fig}[1]{Figure \ref{#1}}
\newcommand{\Figg}[2]{Figures \ref{#1} and \ref{#2}}
\newcommand{\Figc}[2]{Figure \ref{#1}\thinspace{}\fc{#2}}
\newcommand{\Figcc}[3]{Figure \ref{#1}\thinspace{}\fc{#2} and \fc{#3}}

\renewcommand{\l}{\left(}
\renewcommand{\r}{\right)}
\newcommand{\T}{\mathcal{T}}
\newcommand{\gs}{\text{gs}}

\newcommand{\bra}[1]{\langle#1|}
\newcommand{\bkt}[2]{\left\langle #1 |#2 \right\rangle}
\renewcommand{\ij}{{\langle i, j \rangle}}
\renewcommand{\H}{\hat{\mathcal{H}}}
\newcommand{\Ht}{\tilde{\mathcal{H}}}
\renewcommand{\c}{\hat{c}}
\newcommand{\f}{\hat{f}}
\newcommand{\tf}{\hat{\tilde{f}}}
\newcommand{\tfd}{\hat{\tilde{f}}^\dagger}
\renewcommand{\a}{\hat{a}}
\newcommand{\cd}{\hat{c}^\dagger}
\newcommand{\rh}{\hat{\rho}}
\newcommand{\rht}{\tilde{\rho}}
\newcommand{\ad}{\hat{a}^\dagger}
\newcommand{\bd}{\hat{b}^\dagger}
\newcommand{\ubd}{\hat{\uline{b}}^\dagger}
\newcommand{\ub}{\hat{\uline{b}}}
\renewcommand{\b}{\hat{b}}
\newcommand{\hd}{\hat{h}^\dagger}
\newcommand{\h}{\hat{h}}
\newcommand{\dd}{\hat{d}^\dagger}
\renewcommand{\d}{\hat{d}}
\newcommand{\n}{\hat{n}}
\newcommand{\D}{\hat{D}}
\newcommand{\Dd}{\hat{D}^\dagger~\hspace{-0.12cm}}

\newcommand{\G}{\hat{\Gamma}}
\newcommand{\Gd}{\hat{\Gamma}^\dagger}
\newcommand{\F}{\hat{F}}
\newcommand{\Fd}{\hat{F}^\dagger}
\newcommand{\hc}{\text{h.c.}}
\newcommand{\MF}{\text{MF}}
\newcommand{\BEC}{\text{BEC}}
\newcommand{\RG}{\text{RG}}
\newcommand{\psd}{\hat{\psi}^\dagger}
\newcommand{\ps}{\hat{\psi}}
\newcommand{\I}{\text{I}}
\newcommand{\p}{\text{p}}
\newcommand{\fd}{\hat{f}^\dagger}
\newcommand{\s}{\text{S}}
\renewcommand{\sf}{\text{MIX}}
\renewcommand{\O}{\hat{\mathcal{O}}}
\newcommand{\U}{\hat{U}}
\newcommand{\W}{\hat{W}}
\newcommand{\Ud}{\hat{U}^\dagger}
\newcommand{\KP}{\text{KP}}
\newcommand{\HMF}{\mathscr{H}_{\text{MF}}}
\newcommand{\ph}{\text{ph}}
\newcommand{\IB}{\text{IB}}
\newcommand{\B}{\text{B}}
\newcommand{\eff}{\text{eff}}

\renewcommand{\d}{\ket{\downarrow}}
\renewcommand{\u}{\ket{\uparrow}}
\newcommand{\g}{\ket{g}}
\newcommand{\psk}[1]{\ket{\psi(#1)}}
\newcommand{\phk}[1]{\ket{\phi(#1)}}
\newcommand{\psb}[1]{\bra{\psi(#1)}}
\newcommand{\phb}[1]{\bra{\phi(#1)}}
\newcommand{\mun}{\delta \mu_n^\text{ec}}
\newcommand{\mup}{\delta \mu_p^\text{ec}}
\newcommand{\RR}{\ve R}
\newcommand{\CC}{\mathcal{C}}

\newcommand{\diff}{\mathop{}\!\mathrm{d}}

\newcommand{\Shat}{\hat{\mathbf{S}}}
\newcommand{\shat}{\hat{S}}

\renewcommand{\sp}{\ket{\text{sp}(\ve{Q})}}
\newcommand{\KC}{K${}_3$C${}_{60}$ }

\def\bra#1{\mathinner{\langle{#1}|}}
\def\ket#1{\mathinner{|{#1}\rangle}}
\def\braket#1{\mathinner{\langle{#1}\rangle}}
\def\Bra#1{\left<#1\right|}
\def\Ket#1{\left|#1\right>}

\newenvironment{formal}{%
  \def\FrameCommand{%
    \hspace{1pt}%
    {\color{blue}\vrule width 2pt}%
    {\color{formalshade}\vrule width 4pt}%
    \colorbox{formalshade}%
  }%
  \MakeFramed{\advance\hsize-\width\FrameRestore}%
  \noindent\hspace{-4.55pt}
  \begin{adjustwidth}{}{7pt}%
  \vspace{2pt}\vspace{2pt}%
}
{%
  \vspace{2pt}\end{adjustwidth}\endMakeFramed%
}

\newtheorem{defn}{Definition}
\newtheorem{corr}{Corrolary}
\newtheorem{lemma}{Lemma}
\newtheorem{thm}{Theorem}
\newtheorem{remark}{Remark}
\newtheorem{notation}{Notation}

\sloppy

\newpage
\section*{Supplementary material}

\subsubsection*{\textbf{\emph{Preparation of Fermi-Hubbard chains}}}

Our experiments start with a degenerate spin-balanced mixture of fermionic $^{6}$Li atoms in the lowest hyperfine states ($F=1/2;m_{F}=\pm 1/2$). 
This ultracold gas is loaded into a single 2D plane of a $27\,E^{z}_{\text{r}}$ vertical lattice with spacing $3.1\,\mu$m, where $E^{l}_{\text{r}} = h^{2}/8md^{2}_{l}$ is the recoil energy, $m$ is the atomic mass, and $d_{l}$ is the lattice spacing along direction $l$.
The total atom number in the plane is tuned by varying the parameters of our evaporative cooling sequence.
The cloud is then converted to independent 1D chains by ramping up the $y$- lattice, with a spacing of $2.3\,\mu$m, from $0\,E^{y}_{\text{r}}$ to $27\,E_{\text{r}}^{y}$ in $100\,$ms.
For experiments performed at $U/t = 15$, the $x$-lattice, with a spacing of $1.15\,\mu$m, is simultaneously ramped up from $0\,E^{x}_{\text{r}}$ to $7\,E_{\text{r}}^{x}$. 
The tunneling rates for the $x$- and $y$-directions are extracted from band width calculations to be $250\,$Hz and $1.3\,$Hz respectively. 
During the lattice ramp, the scattering length is tuned from $230\,a_{B}$ to $2150\,a_{B}$. 
The onsite interaction is $U/h = 3.75\,$kHz, as calculated from the ground band Wannier functions. 
The corresponding final spin exchange amplitude is $J = 4t^2/U = h \times 65\,$Hz. 

At the end of the adiabatic ramps, the local quench is performed by a $20\,\mu$s pulse of the focused pushout beam and the system is let to evolve for different periods of time. 
For detection, the chains are frozen in place by quickly ramping up the $x$-lattice to $33\,E^{x}_{\text{r}}$, suppressing any further dynamical evolution. 
Our standard Stern-Gerlach spin-detection technique is then performed using a superlattice~\cite{Boll2016}.
Finally, fluorescence images are taken via Raman sideband cooling in a pinning lattice, enabling single-site spin and density resolution with a fidelity of $97\%$~\cite{Boll2016,Omran2015}. 

The temperature in our system is estimated to be $0.75\pm\,0.03 J$ by comparison of the experimental mean nearest and next-nearest neighbour spin correlations without the quench to exact diagonalization predictions of the Heisenberg model.

Fig. 3B and Fig. 3C include data taken at $U/t = 8, 11$ and $20$.
This is achieved by keeping $U$ fixed and tuning the final ramp values of the $x$-lattice to achieve $t/h = 410\,$Hz, $320\,$Hz and $190\,$Hz.
The corresponding spin exchange amplitudes are $J/h = 207\,$Hz, $115\,$Hz and $38\,$Hz, respectively.

\renewcommand{\thefigure}{S\arabic{figure}}
\setcounter{figure}{0}

\begin{figure*}[t!]
  \centering
    \includegraphics[width=0.65\textwidth]{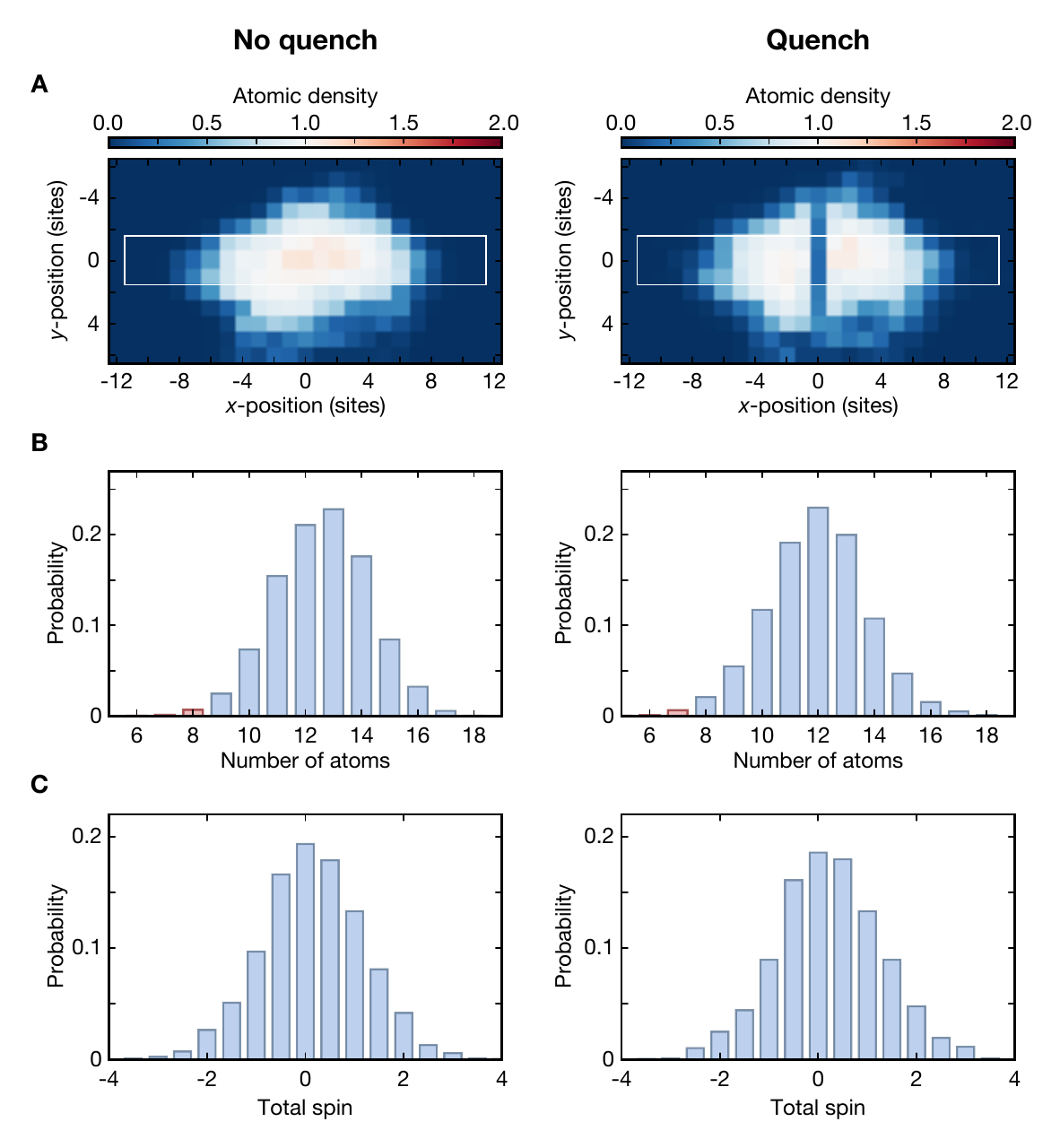}
    \caption{
\textbf{Chain properties with and without the quench.
} \textbf{A,}  Density profile of the cloud without (left) and immediately after (right) the quench. 
The white boxes indictes the central three Hubbard chains which are used in the analysis. 
\textbf{B,} Atom number distribution per chain without (left) and after (right) the quench. Red bars indicate data which was excluded from the analysis.
\textbf{C,} Magnetization $\Sigma_i \hat{S_i^z}$ of the analyzed chains without (left) and after (right) the quench.
The frequency of the pushout beam is tuned to minimize the change in magnetization with and without the quench.  
}
\end{figure*}

\subsubsection*{\textbf{\emph{Pushout beam calibration}}}

Since we prepare multiple 1D chains in one experimental sequence, we use a pushout beam with a high aspect ratio to simultaneously quench the central site of multiple chains of atoms in the lattice gas. 
We tune the frequency of the beam to realize spin-independent removal of the atom (Fig. S1C). 

The focus and power of the beam are calibrated to maximize the pushout efficiency following the quench ${\langle\hat{n}^{\text{h}}_{i}\rangle/(\langle\hat{n}^{\text{h}}_{i-1}\rangle+\langle\hat{n}^{\text{h}}_{i+1}\rangle)}$, where $\langle\hat{n}^{\text{h}}_{i}\rangle$ is the hole density at the addressed site $i$.
This ratio is monitored periodically throughout the data taking process and is kept to a value $\geq\,2$ for the central 3 chains (white box, Fig. S2A and S2D). 
On average, for the dataset at $U/t = 15$, $\langle\hat{n}^{\text{h}}_{i}\rangle \sim 78\%$ and $\langle\hat{n}^{\text{h}}_{i-1}\rangle \simeq \langle\hat{n}^{\text{h}}_{i+1}\rangle \sim 14\%$, which includes contributions of $\sim 4\%$ due to doublon-hole fluctuations. 
However, in the analysis, we post-select on chains which had only a single hole after the quench in the central 9 sites.
This leads to $\langle\hat{n}^{\text{h}}_{i}\rangle\,\sim 81\%$ and reduces $\langle\hat{n}^{\text{h}}_{i-1}\rangle$ and $\langle\hat{n}^{\text{h}}_{i+1}\rangle$ to $\sim 5\%$. 
The probability to have a single hole in the chain immediately after the quench in the analyzed dataset is thus $\sim 91\%$.

\begin{figure*}[t!]
  \centering
    \includegraphics[width=0.75\textwidth]{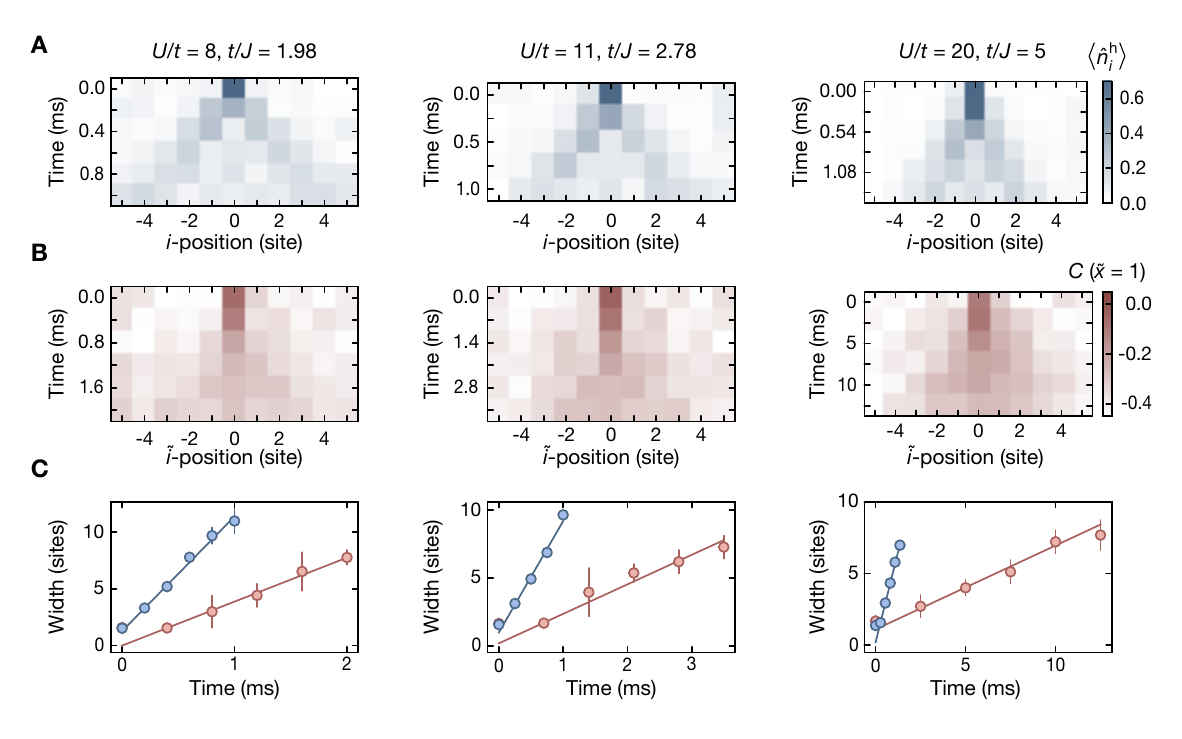}
        \caption{
\textbf{Spin and charge dynamics at different lattice depths.} 
\textbf{A,} The time-resolved hole density $\langle \hat{n}^h_{i}\rangle$ distributions (top), \textbf{B,} nearest neighbour spin correlation $C(\tilde{x}=1)$ distributions and \textbf{C,} their spread as a function of time at $U/t = 8$ (left), $U/t = 11$ (middle) and $U/t = 20$ (right). Slope of solid blue (red) lines gives the velocity of the holon (spinon). Error bars denote 1 s.e.m.
}
\end{figure*}

\subsubsection*{\textbf{\emph{Data analysis}}}

The data used in our results comprise of datasets taken at $U/t\sim8$ (15497 shots), $U/t\sim11$ (11806 shots), $U/t\sim15$ (13578 shots) and $U/t\sim20$ (9848 shots). 
Using a quantum gas microscope with single-site density and spin resolution enables us to determine the total atom number and magnetization of every individual chain. 
Only clouds with a total atom number between $60$ and $90$ are chosen to ensure unity filling of the central three chains of the cloud.
To avoid averaging results over varying density profiles due to the harmonic trap, we analyze the out of equilibrium dynamics of only these central three chains.
The analysis is also done only for chains which had $\geq 9$ atoms in the absence of the quench and $\geq 8$ atoms after the quench. 
Some characteristic statistics for the analyzed chains are shown in Fig. S1 for $U/t = 15$, with and without the quench.

\subsubsection*{\textbf{\emph{Velocity extraction at different $U/t$}}}

The velocity extraction procedure and analysis is discussed in the main text for $U/t = 15$, with ${t/h = 250\,}$Hz and $J/h = 65\,$Hz. 
The velocities of the charge and spin excitations at different lattice depths corresponding to $U/t = 8, 11$ and $20$ were extracted using an identical analysis.
First, the full width of the hole density $\langle\hat{n}^{\text{h}}_{i}\rangle$ and squeezed space nearest neighbour correlation $C(\tilde{x}= 1)$ distributions at 30\% of maxima are calculated at different times after the quench (see Fig. S2). 
For the squeezed space analysis, sites with holes or doublons are removed from the central 9 sites, except for nearest neighbour doublon-hole pairs.
The lattice indices are then shifted to the left.  
The resulting widths of the distributions are plotted as a function of time to which a linear function is fit, giving the velocity.
As in the main text, the widths of the distributions immediately after the quench are excluded from the fit. 
The ratio of the extracted velocities of the spin and charge excitations $v_{J}/v_{t}$ at the different lattice depths are found to increase linearly with $J/t$ (see Fig S3), in good agreement with theory.

\begin{figure}[h]
  \centering
    \includegraphics[width=0.5\textwidth]{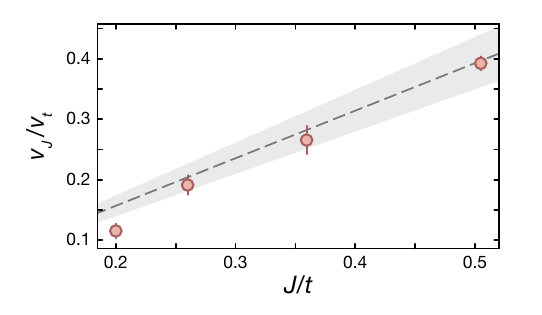}
            \caption{
\textbf{Ratio of quasiparticle velocities as a function of $J/t$.} The extracted ratio of velocities at $J/t = 0.2, 0.26, 0.36$ and $0.5$ corresponding to $U/t = 20, 15, 11$ and $8$ respectively increase linearly with $J/t$. Grey dashed line shows the dependence of the ratio $v^{max}_J/v^{max}_t$ on $J/t$. Grey shade indicates an uncertainty in the estimation of $t$ of $5\%$. Error bars denote 1 s.e.m.
}
\end{figure}

\subsubsection*{\textbf{\emph{Mapping longer range spin correlations}}}

To spatially probe the dynamics of the spin excitation initiated by the quench beyond the nearest neighbour correlations, we map out the measured $\langle\,\hat{S}^{z}_{\tilde{i}}\,\hat{S}^{z}_{\tilde{j}}\,\rangle$ after the quench in squeezed space across the chain (see Fig. S4A), where $\hat{S}^z_{\tilde{i}}$ includes contributions from both the background fluctuations $\hat{S}^z_{\tilde{i},\text{bg}}$ and the spin excitations created by the quench $\hat{S}^z_{\tilde{i},\text{sp}}$.

\begin{figure}[h]
  \centering
    \includegraphics[width=0.5\textwidth]{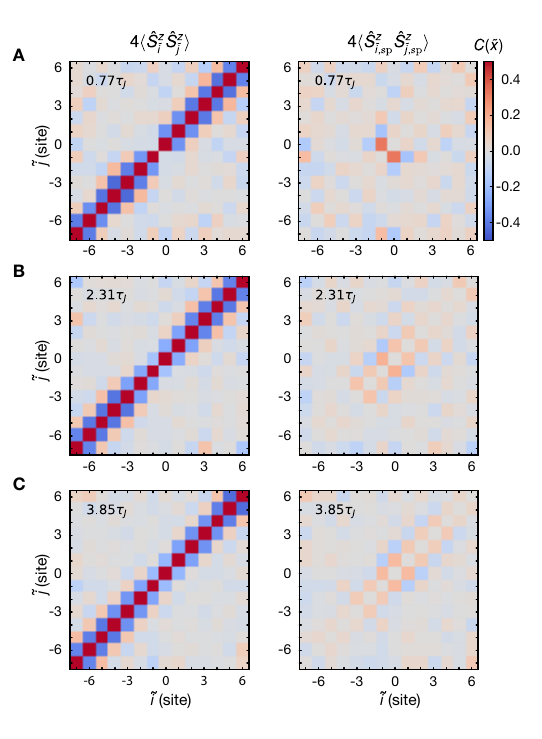}
    \caption{
\textbf{Spin correlations beyond nearest neighbour} The spin correlations $4\langle\,\hat{S}^{z}_{\tilde{i}}\,\hat{S}^{z}_{\tilde{j}}\,\rangle$ before (left column) and after (right column) subtracting the background correlations at different times after the quench. 
The background-subtracted correlations show both the extent of the antiferromagnetic correlations in our chain and the spread of the spin excitation at $0.77\,\tau_J$ (\textbf{A}),$2.31\,\tau_J$ (\textbf{B}) and $3.85\,\tau_J$ (\textbf{C}).
}
\end{figure}

Subtracting the observed signal from the background $\langle\hat{S}^z_{\tilde{i},\text{sp}}\hat{S}^z_{\tilde{j},\text{sp}}\rangle = \langle\hat{S}^z_{\tilde{i}}\hat{S}^z_{\tilde{j}}\rangle - \langle\hat{S}^z_{\tilde{i},\text{bg}}\hat{S}^z_{\tilde{j},\text{bg}}\rangle$ (see Fig. S4B) enables us to visualize the spatial extent of the spin excitations. One sees a sign reversal of spin correlations even beyond the nearest neighbour.
Such a sign reversal of antiferromagnetic correlations caused when crossing a spinon has been observed in an equilibrium setting~\cite{Salomon2019} and is indicative of the nonlocal nature of the spin excitation.

\subsubsection*{\textbf{\emph{Overlap of quasiparticle distributions}}}

In the main text, the spin-hole-spin correlator ($C_{\text{SHS}}$) was used to demonstrate the spatial separation of the spin and charge excitations. 
The reduction in antiferromagnetic correlations across the hole at longer evolution times than $4\tau_t$ were attributed to the harmonic confinement of our trap which caused the hole to oscillate, changing the overlap of the nearest neighbour spin correlation and hole density distributions.
Fig. S5A shows the hole distribution at longer evolution times than plotted in Fig. 2.
Immediately after the quench, the spin and charge distributions have the maximum overlap (see Fig. S5B).

\begin{figure}[h]
  \centering
    \includegraphics[width=0.5\textwidth]{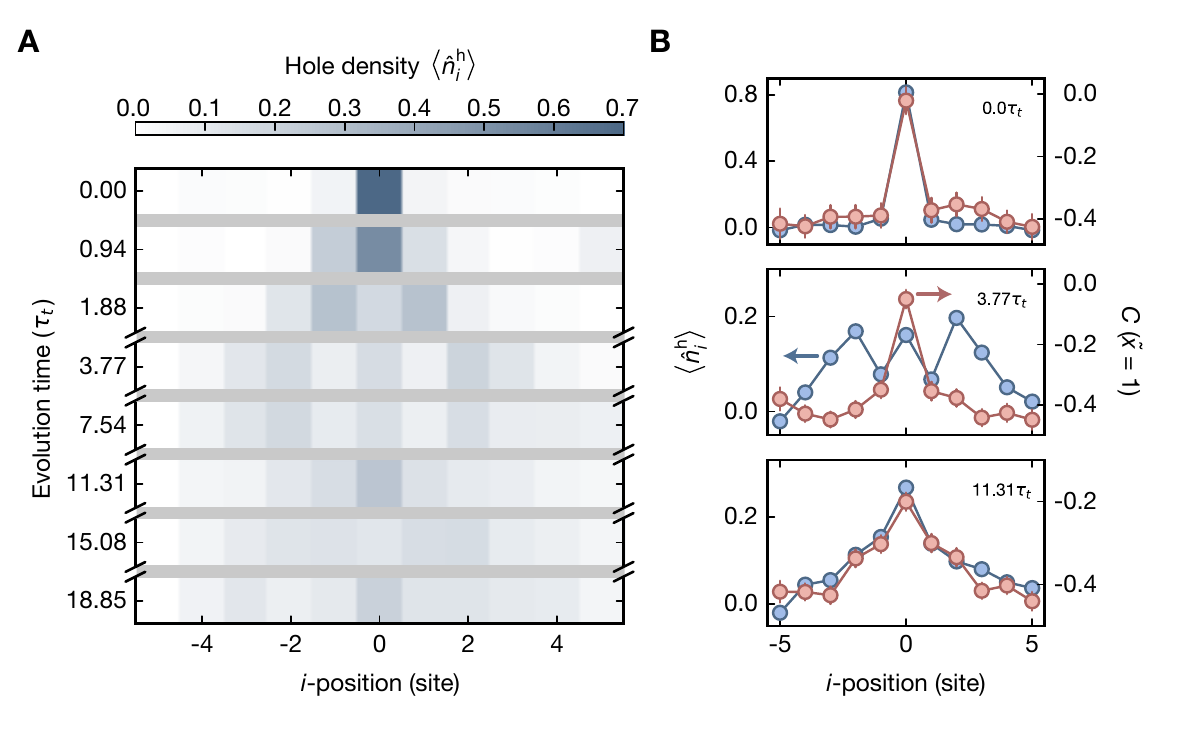}
    \caption{
\textbf{Effect of harmonic confinement on hole dynamics} \textbf{A,} Hole density $\langle \hat{n}^h_{i}\rangle$ distribution at longer evolution times. 
    The hole distribution spreads out and then oscillates back to the center of the chain after $11.31\,\tau_{t}$ and later again after $18.85\,\tau_{t}$ owing to the harmonic confinement in our system.
    \textbf{B,} The oscillatory behaviour changes the overlap of the hole distribution (blue) with the nearest neighbour spin correlation $C(\tilde{x}=1)$ distribution (red). The arrows indicate the respective $y-\,$coordinates. Error bars denote 1 s.e.m.
}
\end{figure}

By $4\,\tau_{t}$, the distributions are maximally separated and at $11.31\,\tau_{t}$, when the hole oscillates back to the center of the chain, the distributions partially overlap again, as the spinon has not spread far from the initial position.

\begin{figure*}[t!]
  \centering
    \includegraphics[width=0.8\textwidth]{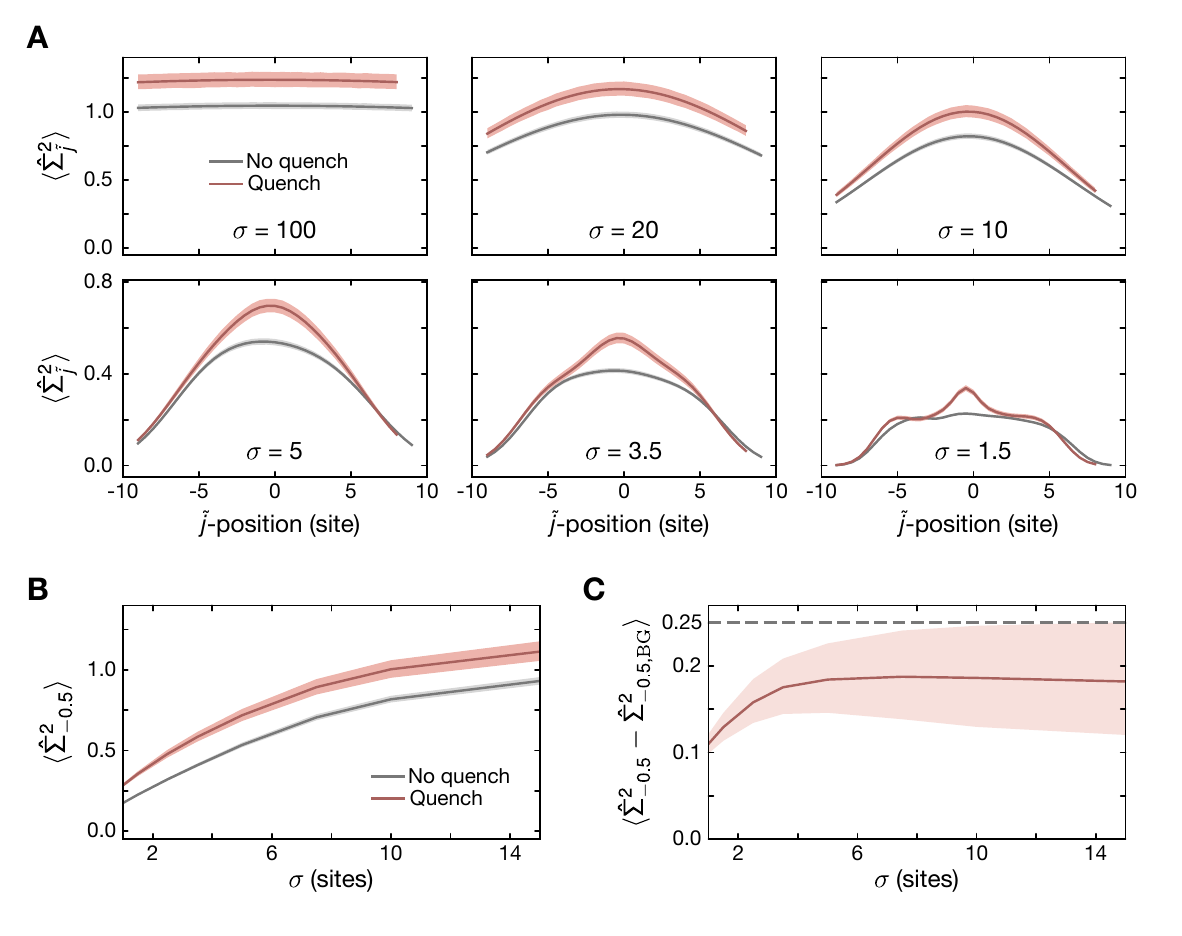}
    \caption{
\textbf{Evaluation of spatially resolved magnetization fluctuations $\hat{\Sigma}^{2}_{\tilde{j}}$ centered at $\tilde{j} = 0$ for different envelope sizes $\sigma$.} \textbf{A}, By comparing datasets with (red) and without (grey) the quench, a large $\sigma$ can be used to extract the total magnetization fluctuation in the chain and a small $\sigma$ can be used to determine the spatial extent of the spin excitation.
    \textbf{B}, The maximum value of $\hat{\Sigma}^{2}$ evaluated at $\tilde{j} = -0.5$ as a function of $\sigma$ with (red) and without (grey) the quench.
    \textbf{C}, Maximum deviation  $\hat{\Sigma}^{2}_{-0.5}-\hat{\Sigma}^{2}_{-0.5,BG}$ obtained by subtracting the two curves in \textbf{B}. At $\sigma$ larger than system size, the measured deviation approaches $0.19\pm0.06$.
Grey and red shades indicate 1 s.e.m. without and after the quench respectively.
}
\end{figure*}

\subsubsection*{\textbf{\emph{Envelope analysis of magnetization fluctuations}}}

In order to spatially evaluate spin fluctuations after removing an atom, the operator $\hat{\Sigma}^{2}_{\tilde{j}} = (\sum_{\tilde{i}}\hat{S}^{z}_{\tilde{i}}f_{\tilde{j}}^{\sigma}(\tilde{i}))^{2}$, which determines magnetization fluctuations within an envelope of width $\sigma$ for an unpolarized chain, was introduced.
When $\sigma$ is larger than the system size, the operator yields the total magnetization fluctuations of the entire chain (see Fig. S6A).
As $\sigma$ is reduced, the spatial extent of the spin fluctuations can be located with better resolution.
In the main text, a width of $\sigma = 1.5$ sites was used to spatially locate the spin excitation. 

At zero temperature and an envelope spread $\sigma$ larger than system size, the measured spin fluctuation signal from the quench $\langle\,\hat{\Sigma}_{\tilde{j}}^{2}\,\rangle$, is expected to be 0.25 due to a single spinon initially located at the quenched site.
However, from our numerical simulations shown in Fig. 4C, due to thermal spin fluctuations at finite temperatures, even an ideal quench would not always create a spinon initially located at the quenched site. 
At our temperature of $k_{\text{B}}T=0.75\,J$ and with $\sigma = 1.5$ sites, we expect to measure a $\langle\,\hat{\Sigma}_{\tilde{j}}^{2}\,\rangle - \langle\,\hat{\Sigma}^{2}_{\tilde{j}}\,\rangle_{\text{BG}}\,\sim\,0.15$.
Taking into account the imperfect quench which creates a single hole with a probability of $91\%$, our measured value of $0.13\pm 0.01$ is in agreement with the expected value.

To capture the contributions of unpaired spin excitations originating from the finite-temperature quench, a larger envelope spread $\sigma$ can be used (see Fig. S6B and Fig. S6C). 
When $\sigma$ is increased to system size, $\langle\,\hat{\Sigma}_{\tilde{j}}^{2}\,\rangle - \langle\,\hat{\Sigma}^{2}_{\tilde{j}}\,\rangle_{\text{BG}}$ approaches $0.19\pm0.06$, in agreement with the expected value of $0.25$ for the creation of a single localized spinon in the chain, within our experimental uncertainties. 
     
  \begin{figure*}
  \centering
    \includegraphics[width=0.8\textwidth]{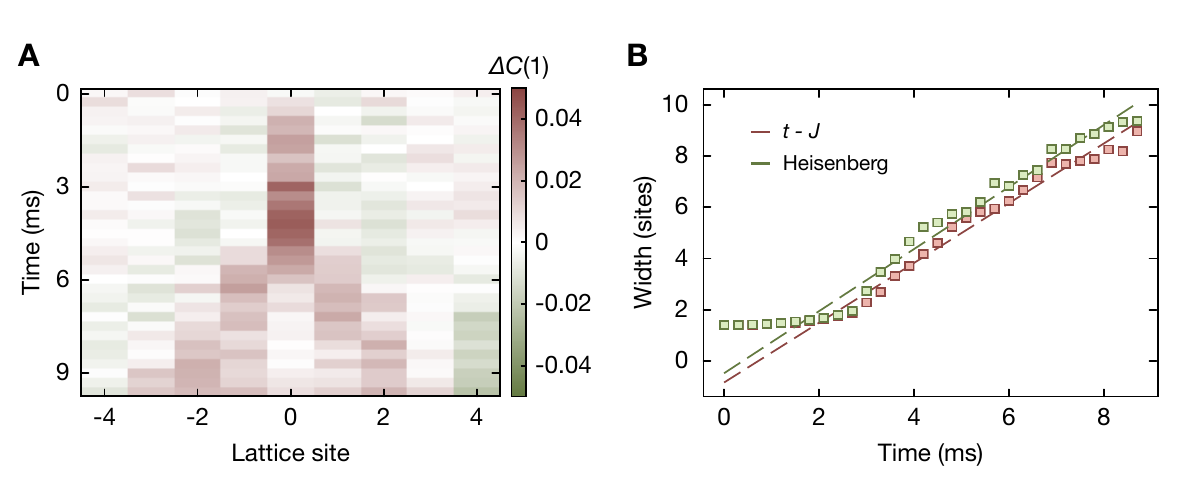}
    \caption{
\textbf{Comparison between the extended $t-J$ and Heisenberg models for $t/J = 3.8$, $T/J = 0.7$ and $V/J=0.456$.} 
    \textbf{A}, Difference between the squeezed space nearest neighbour correlator distribution in the extended $t-J$ model and the Heisenberg model $\Delta C(1)$. 
    \textbf{B}, Velocity extraction for the spinon in the simulation of the extended $t-J$ model (red squares) and the Heisenberg model (green squares). The extracted velocities of $0.58\,\pm\,0.04$ sites/ms for the extended $t-J$ model and $0.61\,\pm\,0.05$ sites/ms for the Heisenberg model are obtained by fitting a straight line to the points (dashed lines).
}
\end{figure*}   
     
\subsubsection*{\textbf{\emph{Extended $t-J$ model simulations}}}

Below half filling, the Fermi-Hubbard model can in leading order in $t/U$ be approximated by the extended $t-J$ model~\cite{Auerbach1994},

\begin{equation}
\begin{split}
\H_{t-J^*} =  \mathcal{P} \biggl[ -t &\sum_{\langle i,j\rangle, \sigma} \hat{c}^{\dagger}_{i,\sigma} \hat{c}_{j,\sigma} + J \sum_{j} \left(\hat{\mathbf{S}}_{j+1} \cdot \hat{\mathbf{S}}_j - \frac{ \hat{n}_{j+1} \hat{n}_j }{4}\right)\\& - \frac{J}{8} \sum_{\langle i,j,r \rangle, \sigma}^{i \neq r}  \biggl( \hat{c}^{\dagger}_{i,\sigma} \hat{c}_{r,\sigma} \n_j - \\& \sum_{\sigma', \tau, \tau'} \hat{c}^{\dagger}_{i,\sigma} \vec{\sigma}_{\sigma,\sigma'} \hat{c}_{r,\sigma'} \cdot  \cd_{j,\tau} \vec{\sigma}_{\tau, \tau'} \hat{c}_{j,\tau'} \biggr) \biggr] \mathcal{P}.
\label{eq:tjmodel}
\end{split}
\end{equation}

Here, $\mathcal{P}$ denotes the projection operator on the subspace without double occupancy, and $\langle i,j,r \rangle$ is a sequence of neighbouring sites. 
The operator $\cd_{j,\sigma}$ creates a fermion with spin $\sigma$ on site $j$ and $\n_{j} = \sum_\sigma \hat{c}^{\dagger}_{j,\sigma} \hat{c}_{j,\sigma}$ is the corresponding density operator. 
The spin operators are defined by $\hat{\textbf{S}}_j = \frac{1}{2} \sum_{\sigma,\sigma'} \hat{c}^{\dagger}_{j,\sigma} \vec{\sigma}_{\sigma,\sigma'} \hat{c}_{j,\sigma'}$, where $\vec{\sigma}$ are the Pauli matrices. 
The first two terms define the $t-J$ model with tunneling of holes with amplitude $t$ and isotropic spin-exchange interactions with coupling constant $J=4 t^2/U$. 
For a single hole, the term $\hat{n}_{j+1}\hat{n}_j$ leads to a constant shift in energy. 
The extended $t-J$ model additionally includes the last term, which describes next-nearest neighbour tunneling of holes correlated with spin-exchange interactions. 

In order to simulate the experiment, we furthermore include a confining harmonic potential for the fermions, given by the additional term
\begin{equation}
\H_{pot} =  V \sum_{i}  (x_i/a)^2 \n_i,
\end{equation}
where $x_i$ is the distance of site $i$ from the center of the chain and $V/J = 0.46$.
\\
In the main text, the experimental results are compared to exact diagonalization calculations of the extended $t-J$ model with a single hole. 
We make use of the conservation of the $z$ component of the total spin and thus obtain a block diagonal Hamiltonian. 
In order to evaluate multi-point correlators as shown in Fig. 4C, we numerically generate snapshots with probabilities given by the time evolved density matrix.

\begin{figure}[t!]
  \centering
    \includegraphics[width=0.5\textwidth]{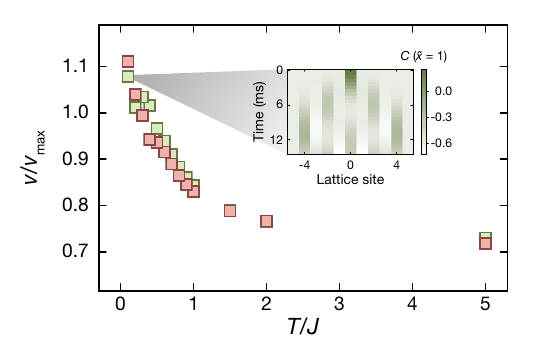}
    \caption{
\textbf{Temperature dependence of the spinon velocity} The ratio of the numerically extracted velocity and the maximum group velocity $v_{\text{max}}^{J} = \pi ^{2} J a_x/h\,$ for $t/J = 3.8$, $V/J=0.456$ and periodic (open) boundary conditions for the spins (hole) as calculated for the extended $t-J$ model (red squares) and the Heisenberg model (green squares). At low temperatures, the coherent motion of the spinon which moves by two sites every step (see inset), makes the extraction of the velocity from a linear fit challenging. 
}
\end{figure}

\subsubsection*{\textbf{\emph{Spinon velocity extraction and comparison to Heisenberg model}}}

The spinon velocity is extracted numerically by measuring the width of the light-cone at 30\% of the maximum value as described in the main text for the experimental results.
For temperatures $T\lesssim J$, the open boundary conditions have a strong effect on the squeezed space nearest neighbour correlator $C(\tilde{x} = 1)$. 
The spin at the boundary is only coupled to one other spin, such that the corresponding correlations are stronger on every second bond in the vicinity of the edge. 
This boundary effect is visible throughout the spin chain in squeezed space and renders the extraction of a spinon velocity with the method described above challenging. 
We therefore simulate periodic boundary conditions for the spins, while the hole is still subject to open boundary conditions and the harmonic potential described above. 
This leads to a smooth behavior of the squeezed space $C(\tilde{x} = 1)$ correlator and therefore enables the extraction of the spinon velocity, which we use in Fig 3A. \\
The spin dynamics in the extended $t-J$ model after the creation of a hole can be directly compared to a Heisenberg spin chain, where initially one site is removed. 
Since in the latter case no hole is involved, the comparison between the two simulations yields insights into the effect of the hole on the spin dynamics. 
In Fig. S7, the relative difference between the squeezed space $C(\tilde{x} = 1)$ correlations for the extended $t-J$ and Heisenberg model are shown. 
Apart from the initial dynamics on the central bond, the relative difference is below 8\% during the entire time evolution, showing how similar the correlations under consideration are in the two simulations. 
As shown in Fig. S8, the extracted spinon velocity decreases with increasing temperature. We attribute this effect to the probability to create a spinon, which exhibits a similar temperature dependence.

\bibliographystyle{unsrt}
\bibliography{main_bib}

\begin{thebibliography}{10}

\bibitem{Giamarchi2004}
T.~Giamarchi.
\newblock {\em {Quantum physics in one dimension}}.
\newblock Clarendon, 2004.

\bibitem{Voit1995}
J.~Voit.
\newblock {One-dimensional Fermi liquids}.
\newblock {\em Rep. Prog. Phys.}, 58:977--1116, 1995.

\bibitem{Kim1996}
C.~Kim, A.~Y. Matsuura, Z.-X. Shen, N.~Motoyama, H.~Eisaki, S.~Uchida,
  T.~Tohyama, and S.~Maekawa.
\newblock {Observation of Spin-Charge Separation in One-Dimensional SrCuO$_2$}.
\newblock {\em Phys. Rev. Lett.}, 77:4054, 1996.

\bibitem{Segovia1999}
P.~Segovia, D.~Purdie, M.~Hengsberger, and Y.~Baer.
\newblock {Observation of Spin and Charge Collective Modes in One-dimensional
  Metallic Chains}.
\newblock {\em Nature}, 402:504--507, 1999.

\bibitem{Kim2006}
B.~J. Kim, H.~Koh, E.~Rotenberg, S.-J. Oh, H.~Eisaki, N.~Motoyama, S.~Uchida,
  T.~Tohyama, S.~Maekawa, Z.-X. Shen, and C.~Kim.
\newblock {Distinct spinon and holon dispersions in photoemission spectral
  functions from one-dimensional SrCuO2}.
\newblock {\em Nat. Phys.}, 2:397--401, 2006.

\bibitem{Auslaender2005}
O.~M. Auslaender, H.~Steinberg, A.~Yacoby, Y.~Tserkovnyak, B.~I. Halperin,
  K.~W. Baldwin, L.~N. Pfeiffer, and K.~W. West.
\newblock {Spin-Charge Separation and Localization in One Dimension}.
\newblock {\em Science}, 308:88--92, 2005.

\bibitem{Jompol2009}
Y.~Jompol, C.~J.~B. Ford, J.~P. Griffiths, I.~Farrer, G.~A.~C. Jones,
  D.~Anderson, D.~A. Ritchie, T.~W. Silk, and A.~J. Schofield.
\newblock {Probing Spin-Charge Separation in a Tomonaga-Luttinger Liquid}.
\newblock {\em Science}, 325:597--601, 2009.

\bibitem{Tserkovnyak2003}
Y.~Tserkovnyak, B.~I. Halperin, O.~M. Auslaender, and A.~Yacoby.
\newblock {Interference and zero-bias anomaly in tunneling between
  Luttinger-liquid wires}.
\newblock {\em Phys. Rev. B}, 68:125312, 2003.

\bibitem{Kinoshita2004}
T.~Kinoshita, T.~Wenger, and D.~S. Weiss.
\newblock Observation of a one-dimensional tonks-girardeau gas.
\newblock {\em Science}, 305:1125--1128, 2004.

\bibitem{Paredes2004}
B.~Paredes, A.~Widera, V.~Murg, O.~Mandel, S.~F{\"o}lling, I.~Cirac, G.~V.
  Shlyapnikov, T.~W. H{\"a}nsch, and I.~Bloch.
\newblock Tonks--girardeau gas of ultracold atoms in an optical lattice.
\newblock {\em Nature}, 429:277, 2004.

\bibitem{Haller2009}
E.~Haller, M.~Gustavsson, M.~J. Mark, J.~G. Danzl, R.~Hart, G.~Pupillo, and
  H.-C. N{\"a}gerl.
\newblock Realization of an excited, strongly correlated quantum gas phase.
\newblock {\em Science}, 325:1224--1227, 2009.

\bibitem{Jacqmin2011}
T.~Jacqmin, J.~Armijo, T.~Berrada, K.~V. Kheruntsyan, and I.~Bouchoule.
\newblock Sub-poissonian fluctuations in a 1d bose gas: from the quantum
  quasicondensate to the strongly interacting regime.
\newblock {\em Phys. Rev. Lett.}, 106:230405, 2011.

\bibitem{Fabbri2015}
N.~Fabbri, M.~Panfil, D.~Cl{\'e}ment, L.~Fallani, M.~Inguscio, C.~Fort, and
  J.-S. Caux.
\newblock Dynamical structure factor of one-dimensional bose gases:
  Experimental signatures of beyond-luttinger-liquid physics.
\newblock {\em Phys. Rev. A}, 91:043617, 2015.

\bibitem{Yang2018}
T.~L. Yang, P.~Gri{\v{s}}ins, Y.~T. Chang, Z.~H. Zhao, C.~Y. Shih,
  T.~Giamarchi, and R.~G. Hulet.
\newblock Measurement of the dynamical structure factor of a 1d interacting
  fermi gas.
\newblock {\em Phys. Rev. Lett.}, 121:103001, 2018.

\bibitem{Boll2016}
M.~Boll, T.~A. Hilker, G.~Salomon, A.~Omran, J.~Nespolo, L.~Pollet, I.~Bloch,
  and C.~Gross.
\newblock {Spin- and density-resolved microscopy of antiferromagnetic
  correlations in Fermi-Hubbard chains}.
\newblock {\em Science}, 353:1257--1260, 2016.

\bibitem{Fiete2007}
G.~A. Fiete.
\newblock {Colloquium: The spin-incoherent Luttinger liquid}.
\newblock {\em Rev. Mod. Phys.}, 79:801--820, 2007.

\bibitem{Hilker2017}
T.~A. Hilker, G.~Salomon, F.~Grusdt, A.~Omran, M.~Boll, E.~Demler, I.~Bloch,
  and C.~Gross.
\newblock {Revealing hidden antiferromagnetic correlations in doped Hubbard
  chains via string correlators}.
\newblock {\em Science}, 357:484--487, 2017.

\bibitem{Salomon2019}
G.~Salomon, J.~Koepsell, J.~Vijayan, T.~A. Hilker, J.~Nespolo, L.~Pollet,
  I.~Bloch, and C.~Gross.
\newblock {Direct observation of incommensurate magnetism in Hubbard chains}.
\newblock {\em Nature}, 565:56--60, 2019.

\bibitem{Cheuk2016}
L.~W. Cheuk, M.~A. Nichols, K.~R. Lawrence, M.~Okan, H.~Zhang, E.~Khatami,
  N.~Trivedi, T.~Paiva, M.~Rigol, and M.~W. Zwierlein.
\newblock {Observation of spatial charge and spin correlations in the 2D
  Fermi-Hubbard model}.
\newblock {\em Science}, 353:1260--1264, 2016.

\bibitem{Parsons2016}
M.~F. Parsons, A.~Mazurenko, C.~S. Chiu, G.~Ji, D.~Greif, and M.~Greiner.
\newblock {Site-resolved measurement of the spin-correlation function in the
  Fermi-Hubbard model}.
\newblock {\em Science}, 353:1253--1256, 2016.

\bibitem{Brown2017}
P.~T. Brown, D.~Mitra, E.~Guardado-Sanchez, P.~Schau{\ss}, S.~S. Kondov,
  E.~Khatami, T.~Paiva, N.~Trivedi, D.~A. Huse, and W.~S. Bakr.
\newblock {Spin-imbalance in a 2D Fermi-Hubbard system}.
\newblock {\em Science}, 357:1385–1388, 2017.

\bibitem{Recati2003}
A.~Recati, P.~O. Fedichev, W.~Zwerger, and P.~Zoller.
\newblock {Spin-Charge Separation in Ultracold Quantum Gases}.
\newblock {\em Phys. Rev. Lett.}, 90:020401, 2003.

\bibitem{Kollath2005}
C.~Kollath, U.~Schollw{\"o}ck, and W.~Zwerger.
\newblock {Spin-Charge Separation in Cold Fermi Gases: A Real Time Analysis}.
\newblock {\em Phys. Rev. Lett.}, 95:176401, 2005.

\bibitem{Kollath2006}
C.~Kollath and U.~Schollw{\"o}ck.
\newblock {Cold Fermi gases: a new perspective on spin-charge separation}.
\newblock {\em New J. Phys.}, 8:220, 2006.

\bibitem{Ogata1990}
M.~Ogata and H.~Shiba.
\newblock {Bethe-ansatz wave function, momentum distribution, and spin
  correlation in the one-dimensional strongly correlated Hubbard model}.
\newblock {\em Phys. Rev. B}, 41:2326--2338, 1990.

\bibitem{Zaanen2001}
J.~Zaanen, O.~Y. Osman, H.~V. Kruis, Z.~Nussinov, and J.~Tworzydlo.
\newblock {The geometric order of stripes and Luttinger liquids}.
\newblock {\em Philos. Mag. B}, 81:1485--1531, 2001.

\bibitem{Castella1995}
H.~Castella, X.~Zotos, and P.~Prelovsek.
\newblock {Integrability and Ideal Conductance at Finite Temperatures}.
\newblock {\em Phys. Rev. Lett.}, 74:972--975, 1995.

\bibitem{Zotos1997}
X.~Zotos, F.~Naef, and P.~Prelovsek.
\newblock {Transport and conservation laws}.
\newblock {\em Phys. Rev. B}, 55:11029--11032, 1997.

\bibitem{Kivelson1982}
S.~Kivelson and J.~R. Schrieffer.
\newblock {Fractional charge, a sharp quantum observable}.
\newblock {\em Phys. Rev. B}, 25:6447--6451, 1982.

\bibitem{Koepsell2018}
J.~Koepsell, J.~Vijayan, P.~Sompet, F.~Grusdt, T.~A. Hilker, E.~Demler,
  G.~Salomon, I.~Bloch, and C.~Gross.
\newblock {Imaging magnetic polarons in the doped Fermi-Hubbard model}.
\newblock {\em arxiv:1811.06907}, 2018.

\bibitem{Omran2015}
Ahmed Omran, Martin Boll, Timon~A. Hilker, Katharina Kleinlein, Guillaume
  Salomon, Immanuel Bloch, and Christian Gross.
\newblock Microscopic observation of pauli blocking in degenerate fermionic
  lattice gases.
\newblock {\em Phys. Rev. Lett.}, 115:263001, 2015.

\bibitem{Auerbach1994}
A.~Auerbach.
\newblock {\em {Interacting Electrons and Quantum Magnetism}}.
\newblock Springer, 1994.

\end{thebibliography}

\end{document}